\documentclass[a4paper,fleqn]{cas-sc}

\usepackage{subfigure}
\usepackage{xcolor}
\usepackage[hyperref=true,backend=biber,sorting=none,backref=true]{biblatex}
\def\tsc#1{\csdef{#1}{\textsc{\lowercase{#1}}\xspace}}
\tsc{WGM}
\tsc{QE}
\begin{document}
\let\WriteBookmarks\relax
\def\floatpagepagefraction{1}
\def\textpagefraction{.001}
\shorttitle{Reconstructing Barrow Holographic Dark Energy in {$f(Q,T)$} Gravity and Cosmic Constraint}    

\shortauthors{Zhang et~al.}  

\title [mode = title]{Reconstructing Barrow Holographic Dark Energy in {$f(Q,T)$} Gravity and Cosmic Constraint}  




\author[1,2]{Xuwei Zhang}
% [orcid=0009-0002-9943-4062]

% \ead{zhangxuwei@xao.ac.cn}

\author[3,1,2,6]{Xiaofeng Yang}
\cormark[1]
% Footnote of the second author
% Email id of the second author
\ead{xfyang@henu.edu.cn}

\author[1,3]{Yunliang Ren}
% [orcid=0000-0002-9659-4581]
% Footnote of the second author
% Email id of the second author
% \ead{}

\author[1,3]{Shuangnan Chen}
% [orcid=0009-0001-5887-7652]
% Footnote of the second author
% Email id of the second author
% \ead{}

\author[1,5]{Yangjun Shi}
% [orcid=0009-0007-8857-7371]
% Footnote of the second author
% Email id of the second author
% \ead{}
\author[1,2]{Cheng Cheng}
\cormark[1]
\ead{chengcheng@xao.ac.cn}

\author[1]{Xiaolong He}

\affiliation[1]{organization={State Key Laboratory of Radio Astronomy and Technology, Xinjiang Astronomical Observatory, Chinese Academy of Sciences},
            % addressline={150 Science 1-Street}, 
            city={Urumqi},
%          citysep={}, % Uncomment if no comma needed between city and postcode
            postcode={830011}, 
            country={China}}
\affiliation[2]{organization={School of Astronomy and Space Science, University of Chinese Academy of Sciences},
            addressline={No.19A Yuquan Road},
            city={Beijing},
            postcode={100049},
            country={China}}
\affiliation[3]{organization={School of Physics and Electronics, Henan University},
            city={Kaifeng},
            postcode={475004},
            country={China}}
\affiliation[4]{organization={School of Physical Science and Technology, Xinjiang University},
            addressline={666 Shenli Street},
            city={Urumqi},
            postcode={830046},
            country={China}}
\affiliation[5]{organization={School of Physics and Astronomy, China West Normal University},
            addressline={No1. Shida Road},
            city={Nanchong},
            postcode={637002},
            country={China}}

% \affiliation[6]{organization={Key Laboratory of Radio Astronomy and Technology, Chinese Academy of Sciences},
%             addressline={A20 Datun Road, Chaoyang District},
%             city={Beijing},
%             postcode={100101},
%             % state={},
%             country={China}}

\affiliation[6]{{Xinjiang Key Laboratory of Radio Astrophysics},
            addressline={150 Science 1-Street},
            city={Urumqi, Xinjiang},
            postcode={830011},
            country={China}}

\cortext[cor1]{Corresponding author}

\begin{abstract}
In this work, we reconstruct the cosmological evolution of Barrow Holographic Dark Energy (BHDE) within the framework of modified gravity $f(Q,T)$. Working in the coincident gauge, we incorporate the holographic principle into non-metric gravity with non-minimal matter coupling. To address the dynamical complexity, we adopt a reconstruction approach, deriving the matter density evolution from the modified field equations rather than imposing a conserved fluid a priori. We perform parameter estimation using the latest observational data, including Type Ia supernovae, BAO, and direct Hubble parameter measurements. Our results show that the model provides a theoretical framework to describe late-time cosmic evolution and the universe's accelerated expansion. Despite the additional complexity introduced, the model offers an alternative approach for investigating dark energy within modified gravity theories.
\end{abstract}


% \begin{highlights}
% \item 
% \item 
% \item 
% \end{highlights}

% Keywords

\maketitle

% Main text
\section{Introduction}\label{sec:intro}

Over the past few decades, a series of major discoveries in cosmology have profoundly reshaped our understanding of the universe. In 1998, observations of Type Ia supernova first revealed that the universe is undergoing accelerated expansion \cite{perlmutter_discovery_1998, Riess_1998}. This groundbreaking discovery was later confirmed by various other cosmological observations, including measurements of temperature anisotropies and polarization in the cosmic microwave background (CMB) radiation \cite{1992ApJ...396L...1S, 2020Planck}, the peak length scale of baryon acoustic oscillations (BAO) \cite{Eisenstein_2005, 10.1111/j.1365-2966.2011.19592.x}, the evolution of the large-scale structure (LSS) of the universe \cite{Dodelson_2002, Percival_2007}, and direct measurements of the Hubble parameter using cosmic chronometers \cite{Daniel_Stern_2010, Moresco_2015}. These observations suggest dark energy (DE) with negative pressure, driving this acceleration against gravitational collapse, though its nature remains elusive.

For such accelerated expansion to occur, dark energy must produce a repulsive gravitational effect that permeates the entire observable universe. Ordinary baryonic matter, however, does not exhibit the properties required to explain this phenomenon, nor can it account for such a significant portion of the universe's energy budget. As a result, researchers have proposed a variety of alternative theories and models to the nature of dark energy and the cosmic acceleration it causes \cite{10.1093/oso/9780198526827.001.0001}.

The simplest and most widely accepted theory is $\Lambda \text{CDM}$ model, where $\Lambda$ denotes cosmological constant introduced by Einstein \cite{Carroll_2001}. Based on $\Lambda \text{CDM}$ model, the latest observations suggest that our universe consists of 68.3\% dark energy, 26.8\% cold dark matter and 4.9\% ordinary matter \cite{2020Planck}. However, it faces challenges: the fine-tuning problem (why $\Lambda$ is so small), the cosmic coincidence issue (why dark energy dominates now), and the Hubble tension-a discrepancy between the CMB-derived $H_0=67.4\pm0.5 \text{km/s/Mpc}$ by Planck \cite{2020Planck} and local SNIa-based $73.04\pm 1.04 \text{km/s/Mpc}$ \cite{Riess_2022}.

An alternative approach modifies general relativity at large scales \cite{Clifton_2012}. Extensive research has explored modifications based on curvature, such as $f(R)$ \cite{1970MNRAS.150....1B} and $f(G)$ gravity \cite{NOJIRI20051,NOJIRI_2007}, as well as those based on torsion, like $f(T)$ gravity \cite{Cai_2016,Bengochea_2009}. Recently, Symmetric Teleparallel Gravity, or $f(Q)$ gravity, has gained attention for describing gravity through non-metricity while keeping curvature and torsion zero \cite{HEISENBERG20241, CAPOZZIELLO2022101113, CAPOZZIELLO2022137229}. Furthermore, theories introducing non-minimal couplings between geometry and matter have been proposed to address cosmic acceleration. Following the proposal of $f(R,T)$ gravity \cite{PhysRevD.84.024020}, various extensions have emerged, including $f(R,G)$ \cite{Bamba2009FinitetimeFS}, $f(T,B)$ \cite{Bahamonde_2015,Bahamonde_2017}, $f(R,T^2)$ \cite{Kat_rc__2014}, and notably $f(Q,T)$ gravity \cite{Xu_2019, De_2024, Harko_2014}. Among these, we focus on $f(Q,T)$ gravity, where the non-minimal coupling implies that the energy-momentum tensor is generally not conserved, leading to an energy exchange between geometry and matter.

The cosmological constant problem represents a fundamental crisis at the intersection of quantum field theory (QFT) and general relativity \cite{martin_everything_2012}. To address this, Holographic Dark Energy (HDE) applies the 't Hooft holographic principle to cosmology \cite{hooft2009dimensionalreductionquantumgravity}. Based on the Cohen et al. UV-IR correspondence \cite{cohen_effective_1999}, the entropy of a system is bounded not by its volume but by its surface area. This imposes a constraint relationship between the ultraviolet cutoff (particle physics scale) and the infrared cutoff (cosmological horizon), providing a quantum gravity motivation for the accelerated expansion without fine-tuning a cosmological constant.
After that, Li et al. proposed that the infrared cut-off relevant to the dark energy is the size of the event horizon \cite{LI20041}. Although Hubble cut-off is a natural thought, but Hsu found it might lead to wrong state equation in standard GR \cite{Hsu_2004}.

However, the holographic principle is considered a fundamental property of quantum gravity. Consequently, if this principle holds true, holographic dark energy should be a universal feature present in any viable theory of gravity, not limited to General Relativity. Therefore, it is essential to test the viability of HDE within the framework of modified gravity theories. Our primary motivation is to explore whether the $f(Q,T)$ gravity, when combined with the HDE model, can provide a consistent description of the universe. Specifically, we incorporate the Barrow Holographic Dark Energy (BHDE) model, which includes quantum-gravitational corrections to the horizon entropy, into the non-metric gravity framework. We hypothesize that this combination, using the most natural Hubble horizon as the IR cutoff, can successfully describe the late-time accelerated expansion and observational data without the inconsistencies found in standard GR.

The application of HDE within modified gravity has been widely investigated. Previous studies have reconstructed or analyzed HDE models in various frameworks, such as $f(R)$ \cite{wu_reconstructing_2008}, $f(G)$ \cite{shaikh_holographic_2020}, $f(T, B)$ \cite{zubair_reconciling_2021}, and $f(\mathcal{G},T)$ gravity \cite{sharif_cosmic_2019, alam_renyi_2023}. Specific to non-metricity, Barrow HDE and other variants have also been explored in $f(Q,T)$ \cite{myrzakulov_barrow_2024} and $f(R,T)$ gravity \cite{singh_statefinder_2016,devi_barrow_2024}. Unlike these works which often focus on theoretical reconstruction, we emphasize observational constraints and the consistency of the model with the latest cosmological data.

In this article, we assume that our universe is described by $ f(Q,T) $ gravity, with HDE as one component of the fluid. In Section \ref{sec:mg}, we briefly introduce non-Riemannian geometry and $f(Q,T)$ gravity. In Section \ref{sec:solution}, we incorporate HDE into the model and derive the solution. Section \ref{sec:data} details the observational constraints, followed by results and evolution analysis in Section \ref{sec:result}. Section \ref{sec:conclusion} concludes findings.

\section{$f(Q,T)$ gravity theory}\label{sec:mg}

In 1918, Weyl proposed an extension of Riemannian geometry, introducing a non-metricity tensor $Q_{\alpha \mu \nu}=\nabla_\alpha g_{\mu \nu}=-w_\alpha g_{\mu \nu}$, which describes how the length of a vector changes during parallel transport where $w_\alpha$ corresponds to electromagnetic potentials \cite{Weyl:1918ib}. This framework, known as Weyl geometry, can be further extended to Weyl-Cartan geometry by incorporating spacetime torsion \cite{Xu_2019}.

In Weyl-Cartan geometry, the general affine connection is decomposed into three independent components: the Christoffel symbol $\hat{\Gamma}^\alpha_{\ \mu \nu}$, the contortion tensor $K^\alpha_{\ \mu \nu}$ and the disformation tensor $L^\alpha_{\ \mu \nu}$, expressed as \cite{J_rv_2018}:
\begin{equation}
    \Gamma^\alpha_{\ \mu \nu}=\hat{\Gamma}^\alpha_{\ \mu \nu}+K^{\alpha}{}_{\mu \nu}+L^\alpha{}_{\mu \nu},
\end{equation}
whereas:
\begin{align}
    \hat{\Gamma}^\alpha{}_{\mu \nu}& =\frac{1}{2}g^{\alpha \beta}(\partial_\mu g_{\beta \nu}+\partial_\nu g_{\beta \mu}-\partial_\beta g_{\mu \nu}), \\
    K^{\alpha }{}_{\mu \nu }& = \frac{1}{2}(T^{\alpha }{}_{\mu \nu }+T_{(\mu}{}^{\alpha }{}_{\nu )})=\frac{1}{2}\left(T^{\alpha }{}_{\mu \nu } + T_{\mu}{}^{\alpha}{}_{\nu} + T_{\nu}{}^{\alpha}{}_{\mu} \right), \\
    L^{\alpha}{}_{\mu\nu}& = \frac{1}{2}(Q^{\alpha}_{}{\mu\nu}-Q_{(\mu}{}^{\alpha }{}_{\nu)}) = \frac{1}{2}\left( Q^{\alpha}_{\mu\nu} - Q_{\mu}{}^{\alpha}{}_{\nu} - Q_{\nu}{}^{\alpha}{}_{\mu} \right),
\end{align}
are the standard Levi-civita connection of metric $g_{\mu \nu}$, contortion and disformation tensors respectively.
In the above definitions, the torsion tensors and the non-metric tensor are introduced as follow:
\begin{align}
T^{\alpha}{}_{\mu\nu} &\equiv 2\Gamma^{\alpha}{}_{[\mu\nu]} =\Gamma^{\alpha}{}_{\mu\nu}-\Gamma^{\alpha}{}_{\nu\mu}
,\\
Q_{\rho \mu\nu} &\equiv \nabla_{\rho} g_{\mu\nu} = \partial_\rho g_{\mu\nu} - \Gamma^\beta{}_{\rho \mu} g_{\beta\nu} - \Gamma^\beta{}_{\rho\nu} g_{\mu\beta}.
\end{align}

To ensure a well-defined formulation and avoid ambiguity regarding the affine connection, we adopt the coincident gauge throughout this work. In this gauge, the affine connection is chosen to vanish globally, i.e., $\Gamma^\alpha_{\mu\nu} = 0$. Consequently, the torsion tensor vanishes ($T^\alpha_{\mu\nu}=0$), and the non-metricity tensor reduces to partial derivatives of the metric, $Q_{\rho \mu\nu} = \partial_\rho g_{\mu\nu}$. This choice simplifies the calculations without loss of generality, as the theory remains covariant under coordinate transformations \cite{HEISENBERG20241}. Furthermore, in the coincident gauge, the field equations derived from the variation with respect to the connection are satisfied as constraints for the standard matter content considered here, allowing us to focus on the metric field equations.

The non-metric tensor has two independent traces, namely $Q_{\mu}=Q_{\mu}{}^{\alpha}{}_{\alpha}$ and $\tilde{Q}^{\mu}=Q_{\alpha}{}^{\mu \alpha}$, which differ by the pair of indices being contracted. So we can get quadratic non-metricity scalar as:
\begin{equation}
    Q=\dfrac{1}{4}Q_{\alpha\beta\mu}Q^{\alpha\beta\mu}-\dfrac{1}{2}Q_{\alpha\beta\mu}Q^{\beta\mu\alpha}-\dfrac{1}{4}Q_{\alpha}Q^{\alpha}+\dfrac{1}{2}Q_{\alpha}\tilde{Q}^\alpha\label{Qscalar} .
\end{equation} 

From the perspective of particle cosmology, the non-minimal coupling between the non-metricity $Q$ and the matter trace $T$ implies that the energy-momentum tensor is no longer conserved ($\nabla_\mu T^{\mu\nu} \neq 0$). This non-conservation can be physically interpreted as an irreversible flow of energy between the gravitational field and the matter sector. Phenomenologically, this mimics quantum particle production processes in a time-varying gravitational background, where the vacuum energy decays into material particles, thereby modifying the cosmic expansion history.

Motivated by these considerations, we consider the general form of the Einstein-Hilbert action for the $f(Q,T)$ gravity using units where $8\pi G=1$:
\begin{equation}
S=\int(\frac{1}{2}f(Q,T)+\mathcal{L}_m) \sqrt{-g}  d^4x ,\label{action}
\end{equation}
where $f$ is an arbitrary function of the non-metricity, $\mathcal{L}_m$ is known as matter Lagrangian, $g=\det (g_{\mu \nu})$ denotes determinant of metric tensor, and $T=g^{\mu \nu}T_{\mu \nu}$ is the trace of the matter-energy-momentum tensor, where $T_{\mu \nu}$ is defined as:
\begin{equation}
    T_{\mu \nu}=-\frac{2}{\sqrt{-g}}\frac{\delta(\sqrt{-g}\mathcal{L}_m)}{\delta g^{\mu \nu}}.
\end{equation}
Varying the action \eqref{action} with respect to the metric tensor $g_{\mu\nu}$ we obtain:
\begin{align}
\delta S&=\int \left(\frac{1}{2} \delta[f(Q,T) \sqrt{-g}]+\delta(\mathcal{L}_m \sqrt{-g})\right)d^4x \\
&= \int \frac{1}{2}\left(-\frac{1}{2}f g_{\mu\nu}\sqrt{-g}\delta g^{\mu\nu}+f_Q \sqrt{-g} \delta Q+f_T \sqrt{-g}\delta T -\frac{1}{2}T_{\mu \nu}\sqrt{-g}\delta g^{\mu\nu}\right)d^4x,
\end{align}
where $f(Q,T)$ is simplified to $f$, and $f_Q=\partial f/\partial Q$, $f_T=\partial f/\partial T$. Finally, we obtain the field equation of the $f(Q,T)$ gravity theory:
\begin{equation}
-\frac{2}{\sqrt{-g}}\nabla_\alpha(f_Q \sqrt{-g}P^\alpha_{\ \ \mu \nu})-\frac{1}{2}f g_{\mu \nu}+f_T(T_{\mu \nu}+\Theta_{\mu \nu})-f_Q(P_{\mu \alpha \beta}Q_\nu^{\ \ \alpha \beta}-2Q^{\alpha \beta}_{\ \ \ \ \mu}P_{\alpha \beta \nu})=T_{\mu \nu},\label{FieldEq}
\end{equation}
where tensor $\Theta_{\mu \nu}$ are defined as $g^{\alpha\beta}\delta T_{\alpha\beta}/{\delta g^{\mu\nu}}$ and $P^{\alpha}_{\mu\nu}$ is the super-potential of the model (detailed discussion found in \cite{Xu_2019}). This equation couples non-metricity and matter, extending GR to include geometric and material interactions.

Assuming that the universe is described by an isotropic, homogeneous and spatially flat Friedmann-Lemaitre-Robertson-Walker (FLRW) spacetime, with the line element expressed as:
\begin{equation}
    ds^2=-N^2(t)dt^2+a^2(t)\delta_{ij} dx^i dx^j\label{FLRW},
\end{equation}
where $a(t)$ is the cosmic scale factor used to define the Hubble expansion rate $H=\dot{a}/a$ and the lapse function $N(t)$ used to define dilation rates $\tilde{T}=\dot{N}/N$ (we set $N(t)=1$ without loss of generality). To derive Friedmann equations describing the cosmological evolution, we assume that the matter content of the universe consists of a perfect fluid, whose energy-momentum tensor is given by $T^{\mu}_\nu=\text{diag}(-\rho,p,p,p)$ and tensor $\Theta^{\mu}_\nu$ is expressed as $\text{diag}(2\rho+p,-p,-p,-p)$. Using the line element \eqref{FLRW} and the field equation \eqref{FieldEq}, the modified Friedmann equations in $f(Q,T)$ gravity are:
\begin{align}
    \rho &=\frac{f}{2}-6f_Q H^2-\frac{2f_T}{1+f_T}(\dot{f}_QH+f_Q \dot{H})\label{Fr1}, \\
    p &=-\frac{f}{2}+6f_Q H^2+2(\dot{f}_QH+f_Q \dot{H}),\label{Fr2} 
\end{align}
 where $\dot{f}_Q=\partial f_Q/\partial t$. In the coincident gauge ($\Gamma^{\alpha}_{\mu\nu}=0$), using Eq. \eqref{Qscalar} and the line element \eqref{FLRW}, the non-metricity scalar simplifies to (detailed derivation in \cite{Xu_2019, lu2019cosmologysymmetricteleparallelgravity}):
\begin{equation}
    Q=6H(t)^2.
\end{equation}
The equation of state (EoS) parameter is given by
\begin{equation}
    w=\frac{p}{\rho}=-1+\frac{4 f_Q H+f_Q \dot{H}}{(1+f_T)(f-12f_QH^2)-4 f_T(\dot{f}_QH+f_Q \dot{H})},
\end{equation}
where $\rho=\rho_m+\rho_\text{de}$ and $ p=p_m+p_\text{de}$ represent contributions from baryonic matter and holographic dark energy, respectively, as we focus on the late universe and neglect radiation.

The effective component parameter can be derived from Eq. \eqref{Fr1}\eqref{Fr2} as:
\begin{align}
    \rho_{\text{eff}}&=3H^2=\frac{f}{4f_Q}-\frac{1}{2f_Q}[(1+f_T)\rho+f_T p] \label{F1},\\
    -p_{\text{eff}}&=2\dot{H}+3H^2=\frac{f}{4f_Q}-\frac{2\dot{f}_Q H}{f_Q}+\frac{1}{2f_Q}[(1+f_T)\rho +(2+f_T)p]. \label{F2}
\end{align}
The two equations describe the total effective energy density \(\rho_{\text{eff}}\) and the total effective pressure \(p_{\text{eff}}\), reflecting the combined effects of all ideal fluid components and the modified gravity. The term $f/({4f_Q})$ represents the contribution from modified gravity, while the latter term accounts for the contributions from various fluid components. However, unlike the standard Friedmann equations, there are correction coefficients \(f_T\) and \(f_Q\) that represent interaction modifications. The effective parameters satisfy the conservation equation, but the individual components do not as:
\begin{equation}
    \dot{\rho}_{\text{eff}}+3H(\rho_{\text{eff}}+p_\text{eff})=0,
\end{equation}
though individual components may not conserve due to energy exchange with the modified geometry. Furthermore, the effective EoS using Eq.\eqref{F1}\eqref{F2} can be written as:
\begin{equation}
    w_{\text{eff}}=\frac{p_{\text{eff}}}{\rho_{\text{eff}}} = \frac{-2\dot{H}-3H^2}{3H^2}= -\frac{f - 8\dot{f}_Q H + 2[(1 + f_T)\rho + (2 + f_T)p]}{f - 2[(1 + f_T)\rho + f_T p]},
\end{equation}
describing the universe’s overall dynamics, with acceleration occurring when $w_\text{eff}<-1/3$.

\section{Cosmic solutions with holographic dark energy}\label{sec:solution}

The holographic principle imposes an upper bound on the entropy of the universe. In the HDE model, the energy density of dark energy is typically expressed as \cite{LI20041}:
\begin{equation}
    \rho_\text{de} = 3c^2 M_p^2 L^{-2},
\end{equation}
where $L$ is the characteristic length scale of the universe, $c$ is a free parameter, and $M_p$ is the reduced Planck mass, set to 1 in natural units ($\hbar=c=1$). The Hubble horizon, defined as $L=H^{-1}$, is the simplest choice, though the particle horizon $L_p$ and future event horizon $L_F$ are also viable alternatives. For the Hubble cutoff, the HDE energy density simplifies to:
\begin{equation}
    \rho_{de}=3c^2 H(t)^2 \label{HDE}
\end{equation}
Another HDE model called Barrow Holographic Dark Energy (BHDE) generalizes holographic entropy that arises from quantum-gravitational effects which deform the black-hole surface by giving it an intricate, fractal form. Its energy density is given by:
\begin{equation}
    \rho_{de}=3c^2 H(t)^{2-\Delta}\label{BHDEdef}.
\end{equation}
Here, the exponent $\Delta$ quantifies the quantum deformation of spacetime structure, potentially arising from spacetime foam or other high-energy discrete features. $\Delta=0$ corresponds to the standard smooth spacetime (Bekenstein-Hawking entropy), while $\Delta=1$ indicates a maximal fractal structure, allowing us to probe the microstructure of the horizon using cosmological data \cite{PhysRevD.102.123525}.

In order to incorporate HDE in the modified framework, we adopt a simple form of $f$:
\begin{equation}
    f(Q,T)=m Q^n+\alpha T,
\end{equation}
where $Q=6H^2$, $T=-\rho+3p$, and $m$, $n$ and $\alpha$ are constants. The partial derivatives are:
\begin{equation}
    f_Q= mn Q^{n-1}=mn 6^{n-1}H^{2n-2},\  f_T=\alpha,\ \dot{f}_Q=2 mn(n-1)6^{n-1}H^{2n-3}\dot{H}.
\end{equation}
We also introduce the deceleration factor, which characterizes the acceleration or deceleration of the late universe depending upon its value, defined as:
\begin{equation}
    q=\frac{d}{dt}\frac{1}{H}-1=-\frac{\ddot{a}a}{\dot{a}^2}=-\frac{\dot{H}}{H^2}-1=(1+z)\frac{1}{H(z)}\frac{dH(z)}{dz}-1.   
\end{equation}

To investigate the cosmological dynamics, one would typically substitute the Barrow HDE density and a conserved matter density into the modified Friedmann equations. As derived in detail in \ref{app:derivation}, this yields a general differential equation governing the Hubble parameter for an arbitrary exponent $n$:
\begin{equation}
    \frac{dH}{dz} = \frac{1+\alpha}{2\alpha m n (2n-1) (6H^2)^{n-1} H (1+z)} \left[ \rho_{de} + \left(1+\frac{\alpha}{2}\right)\rho_m + m\left(n-\frac{1}{2}\right)(6H^2)^n \right].
\end{equation}
Solving this non-linear differential equation analytically for an arbitrary $n$ is intractable. To proceed with an analytical investigation, we focus on the baseline case where $n=1$. We explicitly acknowledge that for $n=1$, the gravitational Lagrangian is linear in $Q$, which makes the background dynamics mathematically analogous to linear $f(R,T)$ gravity or GR with non-minimal coupling. While the distinct non-linear geometric features of $f(Q,T)$ gravity strictly emerge when $n \neq 1$, adopting the $n=1$ baseline allows us to derive exact analytical solutions to isolate and study the effects of the coupling parameter $\alpha$.

Furthermore, regarding the inclusion of matter, we adopt a reconstruction approach. We do not impose a standard conserved matter density as prior, as the non-minimal coupling implies energy exchange between matter and geometry. Instead, the matter density $\rho_m$ is derivedfrom the modified field equations as a dynamic output required to satisfy the system's energy balance. To facilitate this derivation and obtain analytical solutions, we parameterize the dark energy equation of state as a constant $w_{de}$.
\begin{equation}
    w_\text{de}=\frac{p_\text{de}}{\rho_\text{de}}.
\end{equation}
By inverting the modified Friedmann equations \eqref{F1} and \eqref{F2} under the ansatz of a constant dark energy equation of state $w_\text{de}$, we explicitly reconstruct the evolution of the energy densities. The resulting expressions for dark energy and matter, derived as dynamic outputs of the system in terms of the Hubble parameter $H(t)$ and its derivative, are:
\begin{align}
    \rho_\text{de}&= \frac{m (2 n-1) \left(6H(t)^2\right)^{n-1} \left((3 \alpha +2) n H'(t)+3 (\alpha +1) H(t)^2\right)}{\left(2 \alpha ^2+3 \alpha +1\right) w_\text{de}}, \label{RD}\\
    \rho_m&= \frac{m (2 n-1) \left(6H(t)^2\right)^{n-1} \left(n (\alpha  (w_\text{de}-3)-2) H'(t)-3 (\alpha +1) (w_\text{de}+1) H(t)^2\right)}{\left(2 \alpha ^2+3 \alpha +1\right) w_\text{de}}\label{RM}.
\end{align}
It is important to emphasize that above equations determine how the fluid densities must evolve to be consistent with the modified geometry. Where in order to get cosmological solution, there is also a simple relation between $H(t)$ and $H(z)$:
\begin{equation}
    \dot{H}(t)=\frac{d}{dt}H(t)=-\frac{d H(z)}{dz}H(z)(1+z).\label{HR}
\end{equation}
Substituting Eq.~\eqref{BHDEdef} into Eq.\eqref{RD} and using the relation \eqref{HR}, we can obtain a differential equation:
\begin{equation}
    3 c^2 H(z)^{2-\Delta}=\frac{m 6^{n-1} (2 n-1) \left(H(z)^2\right)^{n-1} \left(3 (\alpha +1) H(z)^2-(3 \alpha +2) n (z+1) H(z) H'(z)\right)}{\left(2 \alpha ^2+3 \alpha +1\right) {w_\text{de}}}.
\end{equation}
In principle, $H(z)$ can be derived by solving this equation. However, analytical solutions are challenging for higher-order cases. Thus, we first consider $n=1$ and $\Delta=0$ with the initial condition $H(z=0)=H_0$ which denotes value of the Hubble parameter at present, yielding:
\begin{equation}
    H(z)= H_0 (1+z)^{\frac{3 (1+\alpha) \left(m-c^2 w_\text{de}(1+2\alpha)\right)}{m(2+3\alpha)}}.
\end{equation}
We thus obtain the power-law evolution of the Universe which avoids the big-bang singularity similar to the $f(R,T)$ situation in \cite{singhStatefinderDiagnosisHolographic2016}. The corresponding deceleration parameter is:
\begin{equation}
    q = -1+\frac{3 (1+\alpha) \left(m-c^2 w_\text{de}(1+2\alpha)\right)}{m(2+3\alpha)}.
\end{equation}
Here, the deceleration parameter $q$ is a constant that only depends on the model parameters. By choosing specific parameter values, the model can exhibit either accelerated or decelerated expansion. However, since both the deceleration parameter and the effective EoS are time-independent, this model precludes phase transition between accelerated and decelerated expansion phases. To address this limitation and derive a tighter UV cutoff, we use Eq. \eqref{BHDEdef} and set $\Delta=1$, which leads to the following expression for $H(z)$:
\begin{equation}
    H(z)= H_0 (1+z)^{\frac{3 (\alpha +1)}{3\alpha +2}}-(1+2 \alpha) c^2 w_{de} \left((1+z)^{\frac{3 (\alpha +1)}{3\alpha +2}}-1\right)\frac{1}{m}.
\end{equation}
The deceleration parameter becomes:
\begin{equation}
    q=\frac{(2 \alpha +1) c^2 w_\text{de} \left(3 \alpha +(1+z)^{\frac{3 (\alpha +1)}{3 \alpha +2}}+2\right)-H_0 m (1+z)^{\frac{3 (\alpha +1)}{3 \alpha +2}}}{(3 \alpha +2) \left((2 \alpha +1) c^2 w_\text{de} \left((1+z)^{\frac{3 (\alpha +1)}{3 \alpha +2}}-1\right)-H_0 m (1+z)^{\frac{3 (\alpha +1)}{3 \alpha +2}}\right)}.
\end{equation}
And if we set $\Delta=0.5$, we can also get the solution as follows:
\begin{equation}
\begin{split}
H(z) &= \left( 2 \alpha c^2 w_\text{de} + c^2 w_\text{de} \right. \\
&\quad + (1+z)^{\frac{3 (\alpha + 1)}{6 \alpha + 4}} \left( 2 \left( (2 \alpha + 1)^2 c^4 H_0 \right)^{1/2} m^2 w_\text{de}^2 \right. \\
&\quad \left. \left. + c^4 (2 \alpha w_\text{de} + w_\text{de})^2 + H_0 m^2 \right)^{1/2} \right)^2 \frac{1}{m^2}
\end{split}
\end{equation}
In this case, the deceleration factor and effective EOS is also approaching a constant value, similar to the situation of $\Delta = 0$.
However, in other situations if $n \neq 1$ and $\Delta \neq 1\  \text{or}\  0.5$, higher-order differential equations are difficult to solve analytically, so we can only obtain numerical solutions through complex machine computing. We also consider the case where $\alpha=1$ that reduces it to the minimal matter-energy-momentum tensor coupling.

Finally, to assess the viability of our reconstructed model, we compare it against the standard $\Lambda$CDM model. While $\Lambda$CDM assumes a strictly conserved matter component, our comparison is physically justified by checking whether the reconstructed matter density in our model sufficiently mimics the standard cold dark matter behavior:
\begin{equation}
    H(z)=H_0\sqrt{\Omega_m(1+z)^3+\Omega_\Lambda},
\end{equation}
where $\Omega_m=\rho_{m0}/3H_0^2$ is the matter density parameter, and $ \Omega_\Lambda=\rho_{\text{de}0}/3H_0^2$ is the dark energy density parameter. This model assumes that the universe is composed of matter and a cosmological constant $\Lambda$ with no additional exotic components.

\section{Observational data and methodology}\label{sec:data}

\begin{table}[htbp]
    \centering
    \caption{BAO dataset used in the study. The table includes data from the 6dFGS survey \cite{Beutler_2011}, SDSS survey \cite{PhysRevD.103.083533}, and the latest DESI DR2 BAO data\cite{Abdul_Karim_20251,Abdul_Karim_20252}. The table provides effective redshifts $ z_\text{eff} $, along with measurements of the ratio $ D_{\rm M}/r_{\rm d} $, $ D_{\rm H}/r_{\rm d} $, and $ D_{\rm V}/r_{\rm d} $ respectively.}
    \begin{tabular}{lcccc}
    \hline
    Survey          & $z_\text{eff}$ & $D_{\rm M}/r_{\rm d}$ & $D_{\rm H}/r_{\rm d}$ & $D_{\rm V}/r_{\rm d}$\\
    \hline
    6dFGS		    & $0.106$	& & & $2.98\pm0.13$ \\
    \hline
    SDSS MGS 		& $0.15$	& & & $4.51\pm0.14$ \\
    SDSS DR12		& $0.38$    & $10.27\pm0.15$ & $24.89\pm 0.58$ & \\
    SDSS DR12		& $0.51$    & $13.38\pm0.18$ & $22.43\pm 0.48$ & \\
    SDSS DR16 LRG	& $0.70$    & $17.65\pm0.30$ & $19.78\pm0.46$ & \\
    SDSS DR16 ELG	& $0.85$	& $19.50\pm1.00$ & $19.60\pm2.10$ & \\
    SDSS DR16 QSO	& $1.48$    & $30.21\pm0.79$ & $13.23\pm0.47$ & \\
    SDSS DR16 Ly$\alpha$-Ly$\alpha$& $2.33$     & $37.60\pm1.90$& $8.93\pm0.28$ & \\
    SDSS DR16 Ly$\alpha$-QSO	& $2.33$ & $37.30\pm1.70$& $9.08\pm0.34$ & \\
    \hline
    DESI DR2 BGS       & 0.295 & & & $7.944 \pm 0.075$ \\
    DESI DR2 LRG1      & 0.510 & $13.587 \pm 0.169$ & $21.863 \pm 0.427$ & \\
    DESI DR2 LRG2      & 0.706 & $17.347 \pm 0.180$ & $19.458 \pm 0.332$ & \\
    DESI DR2 LRG3+ELG1 & 0.934 & $21.574 \pm 0.153$ & $17.641 \pm 0.193$ & \\
    DESI DR2 ELG2      & 1.321 & $27.605 \pm 0.320$ & $14.178 \pm 0.217$ & \\
    DESI DR2 QSO       & 1.484 & $30.519 \pm 0.758$ & $12.816 \pm 0.513$ & \\
    DESI DR2 Ly$\alpha$ & 2.330 & $38.988 \pm 0.531$ & $8.632 \pm 0.101$ & \\
    \hline
    \end{tabular}
    \label{tab:baodata}
\end{table}

\begin{figure}
    \centering
    \includegraphics[width=0.8\linewidth]{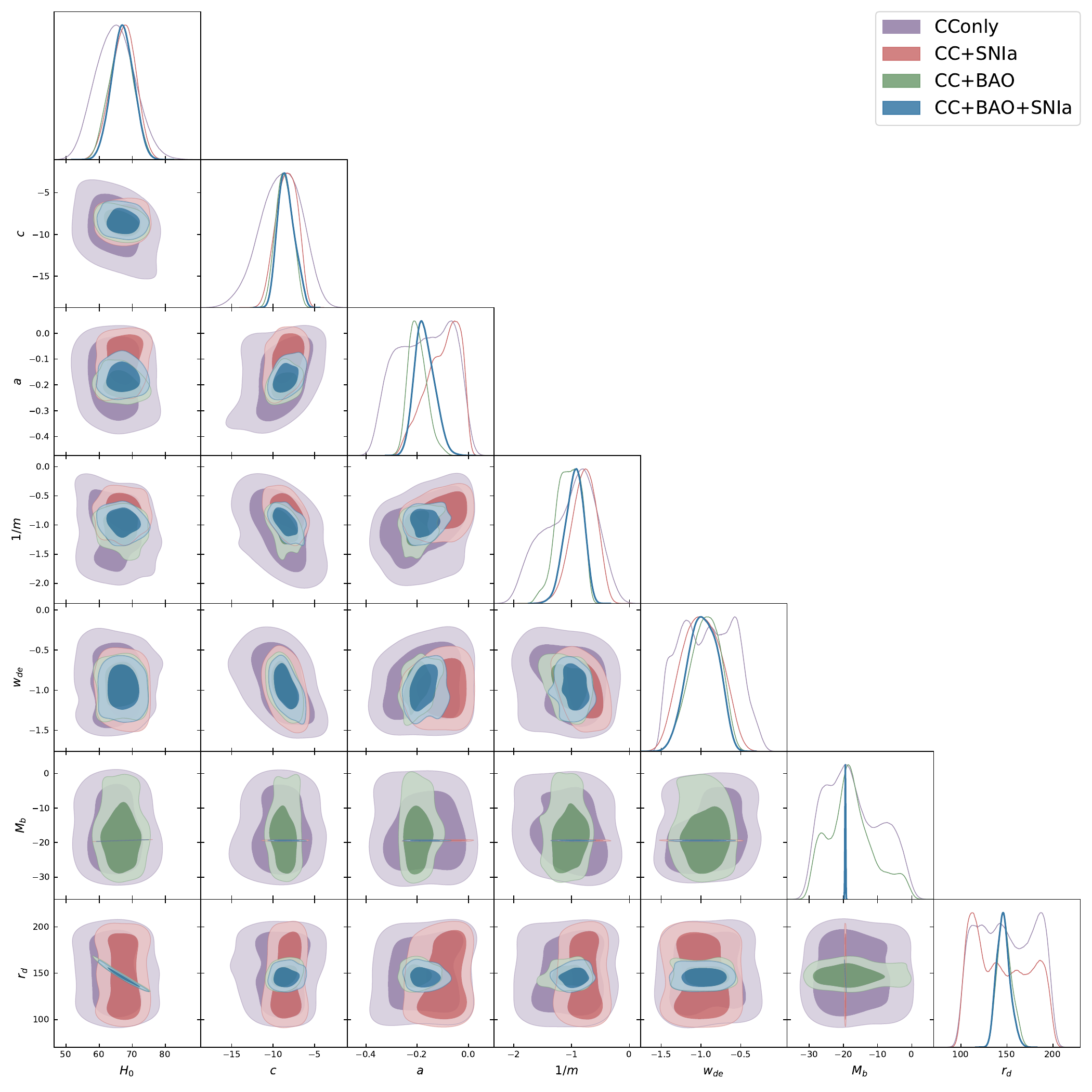}
    \caption{The 1$\sigma$ and 2$\sigma$ confidence contours and the 1D posterior distributions for the HDE model in $f(Q,T)$ gravity with $\Delta=1$ obtained from different dataset combinations. The contours correspond to constraints from CC only (purple), CC+SNIa (red), CC+BAO (green), and the full combination of CC+BAO+SNIa (blue).}
    \label{Fig:constraint}
\end{figure}

In this work, we estimate the cosmological parameters of the model by employing a Markov Chain Monte Carlo (MCMC) method based on the minimization of the chi-square function, $\chi^2$ which is given by \cite{Padilla_2021}:
\begin{equation}
\chi^2 = \sum_i \left(\frac{D_i - T_i(\mathbf{\theta})}{\sigma_i}\right)^2,
\end{equation}
where $D_i$ represents the $i$-th data point, $T_i(\mathbf{\theta})$ is the theoretical prediction for the corresponding quantity, and $\sigma_i$ is the error associated with the $i$-th data point. Here, $\mathbf{\theta}$ denotes the vector of model parameters. To complete the parameter constraints, we utilize the Python package \texttt{emcee}\footnote{\url{https://github.com/dfm/emcee}} \cite{emcee}, a user-friendly MCMC implementation well-suited for cosmological data analysis.

For our analysis, we combine three independent observational datasets:

1. Baryon Acoustic Oscillations (BAO): The BAO measurements provide a standard ruler for distance measurements in the universe. We use the data from the SDSS Baryon Oscillation Spectroscopic Survey (BOSS)  \cite{PhysRevD.103.083533}, Dark Energy Spectroscopic Instrument (DESI) Data Release 2 (DR2) \cite{desicollaboration2024desi2024vicosmological,Abdul_Karim_20251,Abdul_Karim_20252} and 6dF Galaxy Survey (6dFGS) to constrain the cosmological parameters \cite{Beutler_2011}. The comoving horizon distance, the transverse comoving distance
and the volume-averaged distance combining line-of-sight and transverse distances defined as follow:
\begin{align}
    D_H&=\frac{c}{H(z)}, \\
    D_M&=\frac{d_L}{1+z},\\
    D_V&=\left[\frac{cz}{H(z)}\right]^{1/3}\left[\frac{d_L}{1+z}\right]^{2/3},
\end{align}
where $d_L$ is the luminosity distance (defined in Eq.~\eqref{dL}). When scaled by the sound horizon at the drag epoch $r_d$ , ratios such as $D_H/r_d$, $D_M/r_d$, and $D_V/r_d$ serve as important observables for constraining cosmological models and testing the standard model of cosmology.

2. Cosmic chronometers (CC) Measurements: The Hubble parameter measurements, known as the chronometers data, provide independent estimates of the Hubble parameter $H(z)$ at various redshifts. These data serve as an important probe of the expansion rate of the universe. We choose the dataset from \cite{Favale_2023} which includes 32 CC data points shown in Table~\ref{tab:cc_data}. To ensure a rigorous error analysis, we follow the methodology of Moresco et al. \cite{moresco_setting_2020,moresco_unveiling_2022} to construct the full covariance matrix $\mathbf{C}_{\text{CC}}$\footnote{\url{https://gitlab.com/mmoresco/CCcovariance}. This matrix incorporates both statistical uncertainties and systematic effects. The systematic component includes uncertainties related to galaxy stellar-population properties, such as stellar metallicity and potential contamination from young stellar components, which are typically treated as uncorrelated (diagonal). In contrast, the choice of the stellar population synthesis (SPS) model—including the initial mass function (IMF) and the adopted stellar library which induces significant correlated errors across different redshift bins.}

\begin{table}[htbp]
\centering
\caption{Cosmic Chronometers (CC) data compilation used in this work. The Hubble parameter $H(z)$ is given in units of [km s$^{-1}$ Mpc$^{-1}$].}
\label{tab:cc_data}
\begin{tabular}{ccl | ccl | ccl}
\hline
$z$ & $H(z)$ & Ref. & $z$ & $H(z)$ & Ref. & $z$ & $H(z)$ & Ref. \\
\hline
0.070 & $69.0 \pm 19.6$ & \cite{Zhang_2014} & 0.400 & $95.0 \pm 17.0$ & \cite{PhysRevD.71.123001} & 0.875 & $124.5 \pm 17.4$ & \cite{Moresco_2012} \\
0.090 & $69.0 \pm 12.0$ & \cite{Jimenez_2003} & 0.400 & $79.9 \pm 11.4$ & \cite{Moresco_2016} & 0.880 & $90.0 \pm 40.0$ & \cite{Daniel_Stern_2010} \\
0.120 & $68.6 \pm 26.2$ & \cite{Zhang_2014} & 0.425 & $90.4 \pm 12.8$ & \cite{Moresco_2016} & 0.900 & $117.0 \pm 23.0$ & \cite{PhysRevD.71.123001} \\
0.170 & $83.0 \pm 8.0$ & \cite{PhysRevD.71.123001} & 0.450 & $96.3 \pm 14.4$ & \cite{Moresco_2016} & 1.037 & $133.5 \pm 17.6$ & \cite{Moresco_2012} \\
0.179 & $78.0 \pm 6.2$ & \cite{Moresco_2012} & 0.470 & $89.0 \pm 49.6$ & \cite{Ratsimbazafy} & 1.300 & $168.0 \pm 17.0$ & \cite{PhysRevD.71.123001} \\
0.199 & $78.0 \pm 6.9$ & \cite{Moresco_2012} & 0.478 & $83.8 \pm 10.2$ & \cite{Moresco_2016} & 1.363 & $160.0 \pm 33.8$ & \cite{Moresco_2015} \\
0.200 & $72.9 \pm 29.6$ & \cite{Zhang_2014} & 0.480 & $97.0 \pm 62.0$ & \cite{Daniel_Stern_2010} & 1.430 & $177.0 \pm 18.0$ & \cite{PhysRevD.71.123001} \\
0.270 & $77.0 \pm 14.0$ & \cite{PhysRevD.71.123001} & 0.593 & $107.0 \pm 15.5$ & \cite{Moresco_2012} & 1.530 & $140.0 \pm 14.0$ & \cite{PhysRevD.71.123001} \\
0.280 & $88.8 \pm 36.6$ & \cite{Zhang_2014} & 0.680 & $95.0 \pm 10.5$ & \cite{Moresco_2012} & 1.750 & $202.0 \pm 40.0$ & \cite{PhysRevD.71.123001} \\
0.352 & $85.5 \pm 15.7$ & \cite{Moresco_2012} & 0.750 & $98.8 \pm 33.6$ & \cite{Borghi_2022} & 1.965 & $186.5 \pm 50.6$ & \cite{Moresco_2015} \\
0.380 & $86.2 \pm 14.6$ & \cite{Moresco_2016} & 0.781 & $96.5 \pm 12.5$ & \cite{Moresco_2012} & & & \\
\hline
\end{tabular}
\end{table}

3. Type Ia supernova (SNIa) Data: SNIa are considered standard candles because when the light curve reaches its maximum, the absolute luminosity is almost the same. The distance modulus $\mu$ can be obtained according to the following formula:
\begin{equation}
    \mu_{obs}=m-M.
\end{equation}
On the other hand, we can get the theoretical distance modulus from the cosmological model:
\begin{equation}
    \mu_{th}(z)=5\log_{10}d_L(z)+25+M_b,
\end{equation}
where $M_b$ denotes the absolute luminosity of SNIa and the luminosity distance is defined as:
\begin{equation}
    d_L(z)=\frac{c}{H_0}(1+z)\int_0^z \frac{dz'}{E(z')}\label{dL}.
\end{equation}

In this paper, we use Pantheon+ dataset who comprises 1701 SNIa samples \footnote{\url{https://github.com/PantheonPlusSH0ES/DataRelease}}, an increase from the 1048 samples in Pantheon dataset and correspond to light curves of 1550 spectroscopically confirmed SNIa within the redshift range $0.001 < z < 2.26$ \cite{Scolnic_2022,Brout_2022}.

The combined likelihood function $\mathcal{L}$ is then constructed by multiplying the individual likelihoods of each dataset:
\begin{equation}
\mathcal{L} = \mathcal{L}_{\text{BAO}} \times \mathcal{L}_{\text{CC}} \times \mathcal{L}_{\text{SNIa}},
\end{equation}
Thus, the total $\chi^2_\text{tot}$ is:
\begin{equation}
    \chi^2_\text{tot}=\chi^2_{\text{BAO}}+\chi^2_{\text{CC}} +\chi^2_{\text{SNIa}}.
\end{equation}

To test the statistical significance of our constraints, we implement the Akaike Information Criterion (AIC) and Bayesian Information Criterion (BIC), these criteria may help balance model fit and complexity. 
the AIC is given by:
\begin{equation}
\text{AIC} = 2k - 2 \ln(\mathcal{L}),
\end{equation}
where $k$ is the number of parameters and $\mathcal{L}$ is the likelihood. Similarly, the BIC for each model is calculated as:
\begin{equation}
\text{BIC} = k \ln(n) - 2 \ln(\mathcal{L}),
\end{equation}
where $n$ is the number of data points. Models with lower AIC and BIC values are favored, aiding comparisons between HDE variants and $\Lambda$CDM.

\section{Results and analysis}\label{sec:result}

In this research, we use the python package \texttt{GetDist}\footnote{\url{https://github.com/cmbant/getdist}} \cite{Lewis:2019xzd} to visualize the posterior distributions and confidence contours. To ensure the robustness of our results, we performed the analysis using different dataset combinations: Cosmic Chronometers (CC) only, CC combined with Type Ia Supernovae (CC+SNIa), CC combined with Baryon Acoustic Oscillations (CC+BAO), and the full combination of all datasets. The resulting comparisons are shown in Fig.~\ref{Fig:constraint}. We observe a significant overlap among the contours from these different combinations, indicating that the geometrical probes (SNIa and BAO) and the dynamical probe (CC) are consistent within our model framework. Consequently, we focus on the constraints obtained from the full combination, which are summarized in Table \ref{tab:results}.

Here we choose the best fitting model, maximal deformation $f(Q,T)$ HDE with $\Delta=1$, where we find the following 95\% confidence limits for the cosmological and model parameters: the Hubble constant is $ H_0 = 67.1^{+6.5}_{-6.4} \, \text{km/s/Mpc} $, which is consistent with recent Planck measurements, though slightly lower. The parameter $ c $, which characterizes the evolution of dark energy, is constrained to $ c = -8.4^{+2.0}_{-1.7} $. The parameter $ a $, which governs the coupling strength between the energy-momentum tensor and geometry, is found to be $ a = -0.171^{+0.078}_{-0.071} $, indicating that there is a small coupling coefficient between the geometric structure and the fluid term, and the gravitational correction of the interaction is meaningful. The inverse of the coefficient $ 1/m $, reflecting the degree of modification of gravity, is constrained to $ 1/m = -0.95^{+0.30}_{-0.32} $. The EOS parameter for dark energy, $ w_{\text{de}} $, is constrained to $ w_{\text{de}} = -1.01^{+0.38}_{-0.40} $, indicating that dark energy is close to but slightly less than the value for a cosmological constant. The absolute magnitude of the reference galaxy is $M_b = -19.41^{+0.20}_{-0.22}$, with a narrow error range consistent with the expected value for the sample of galaxies considered. Finally, the sound horizon at the drag epoch is measured to be $r_d = 146^{+15}_{-13} \, \text{Mpc}$, in agreement with current BAO measurements.

\begin{figure}
    \centering
    % Left Panel
    \begin{minipage}{0.48\textwidth}
        \centering
        \includegraphics[width=\linewidth]{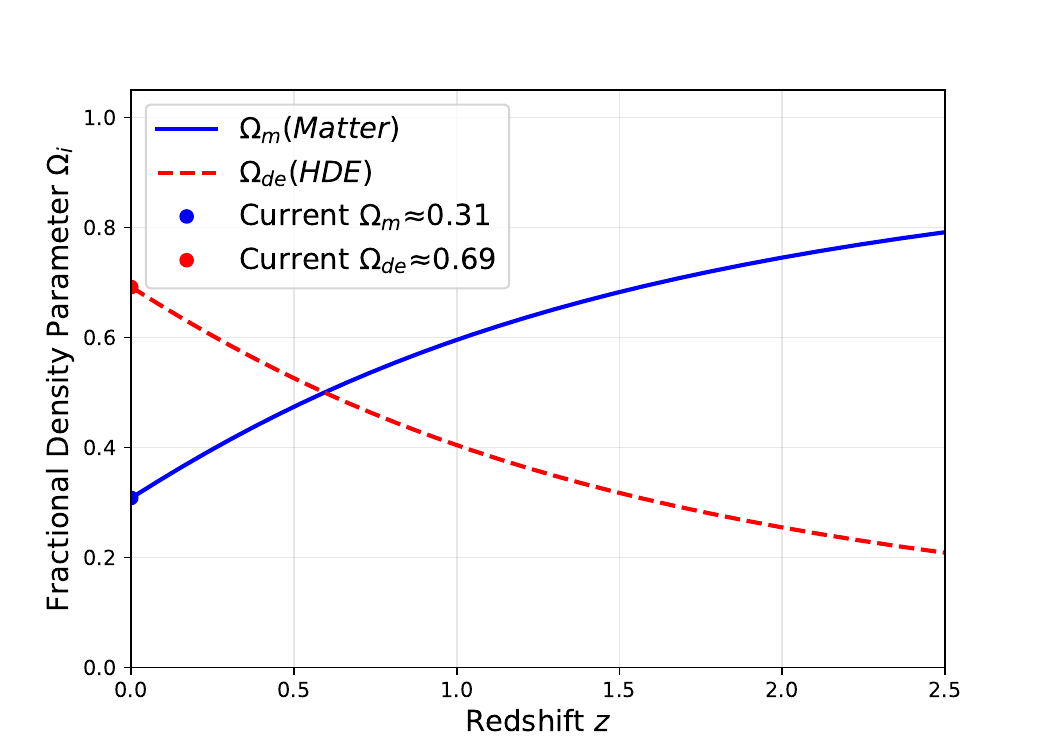}
    \end{minipage}
    \hfill
    % Right Panel
    \begin{minipage}{0.48\textwidth}
        \centering
        \includegraphics[width=\linewidth]{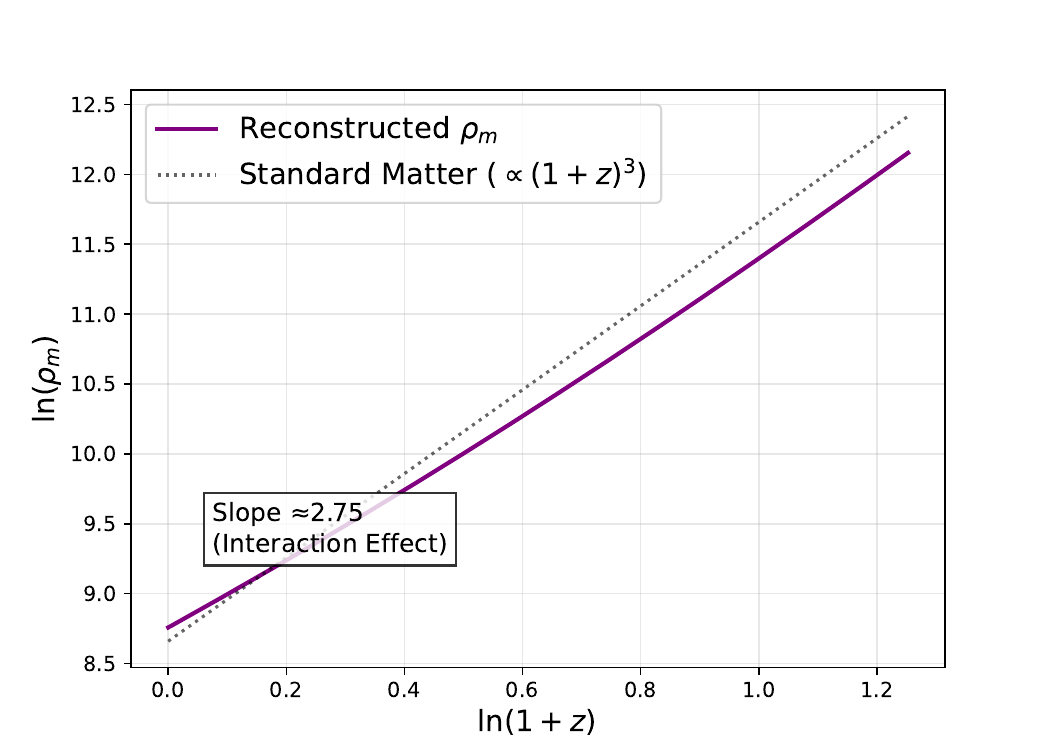}
    \end{minipage}
    
    \caption{\label{Fig:rho_analysis} 
    The evolution of the fractional density parameters $\Omega_m(z)$ (solid blue) and $\Omega_{de}(z)$ (dashed red). The model naturally recovers the matter-dominated era at high redshift and the dark-energy-dominated era at low redshift (left panel). The logarithmic relation between the reconstructed matter density $\rho_m$ and $(1+z)$. The black dotted line represents the standard conservation law ($\rho_m \propto (1+z)^3$).}
\end{figure}

A critical aspect of our analysis is the validation of the matter component, which, as detailed in Section \ref{sec:solution}, is reconstructed from the modified field equations rather than imposed as a conserved fluid. To verify the physical viability of this reconstruction, we plot the evolution of the derived fractional density parameters in Fig.~\ref{Fig:rho_analysis} (left panel) which are inferred from Eq.~(\ref{RD}) and (\ref{RM}). At the present epoch ($z=0$), we find $\Omega_m \approx 0.31$ and $\Omega_{de} \approx 0.69$, which are in agreement with the standard $\Lambda$CDM paradigm and Planck 2018 results. We also analyze the scaling behavior of the reconstructed matter density $\rho_m$ in Fig.~\ref{Fig:rho_analysis} (right panel). The log-log plot reveals a linear relation with a slope of approximately $2.75$. This deviation from the standard dust scaling (slope of 3) is a direct physical consequence of the non-minimal coupling in $f(Q,T)$ gravity, representing the energy exchange between the matter sector and the geometry. This confirms that our model does not lack regular matter; rather, it successfully reconstructs a physically consistent interacting matter component.

\begin{table}[htbp]
    \centering
    \caption{Results of Constraint on parameters with prior ranges and $95\%$ credible limits in HDE $f(Q,T)$ with $\Delta=1$.}
    \begin{tabular}{lcr}
    \hline
        Parameter & Prior & Posterior with 95\% limits\\
        \hline
        {\boldmath$H_0            $} & $[50, 100]$ & $67.1^{+6.5}_{-6.4}$ \\
        
        {\boldmath$c              $} & $[-20, 0]$ & $-8.4^{+2.0}_{-1.7}$ \\
        
        {\boldmath$a              $} & $[-1, 0]$ & $-0.171^{+0.078}_{-0.071}$ \\
        
        {\boldmath$1/m            $} & $[-2, 0]$ & $-0.95^{+0.30}_{-0.32}$ \\
        
        {\boldmath$w_{de}         $} & $[-1.5, 0]$ & $-0.98^{+0.34}_{-0.35}$ \\
        
        {\boldmath$M_b            $} & $[-30, 0]$ & $-19.41^{+0.20}_{-0.22}$ \\
        
        {\boldmath$r_d            $} & $[100, 200]$ & $146^{+15}_{-13}$ \\
        \hline
    \end{tabular}
    \label{tab:results}
\end{table}

\begin{figure}
    \centering
    \includegraphics[width=0.45\textwidth]{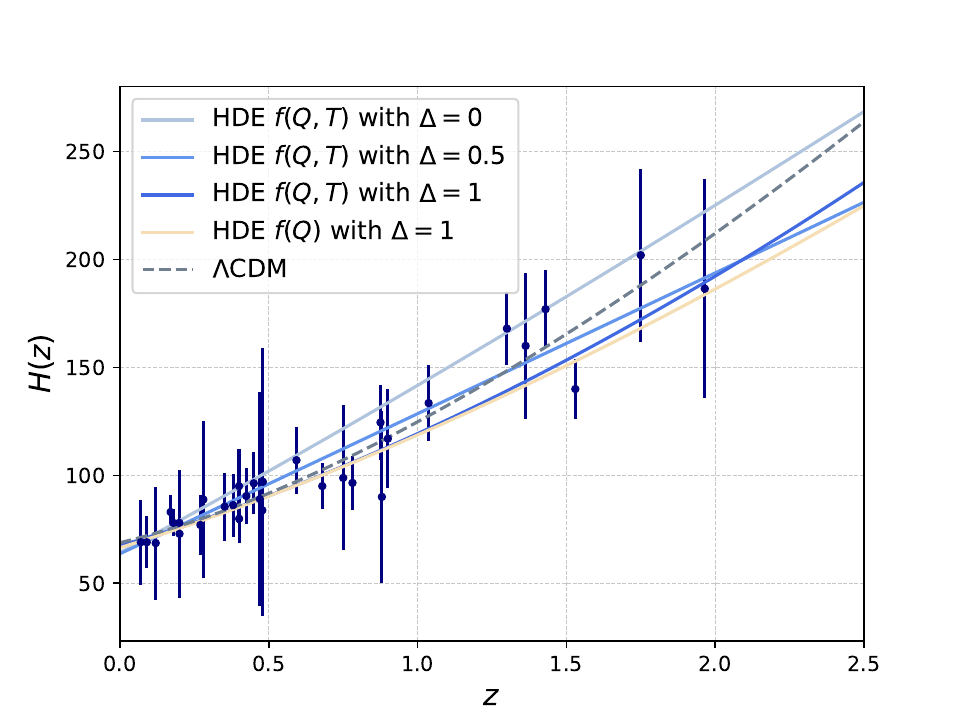}
    \includegraphics[width=0.45\textwidth]{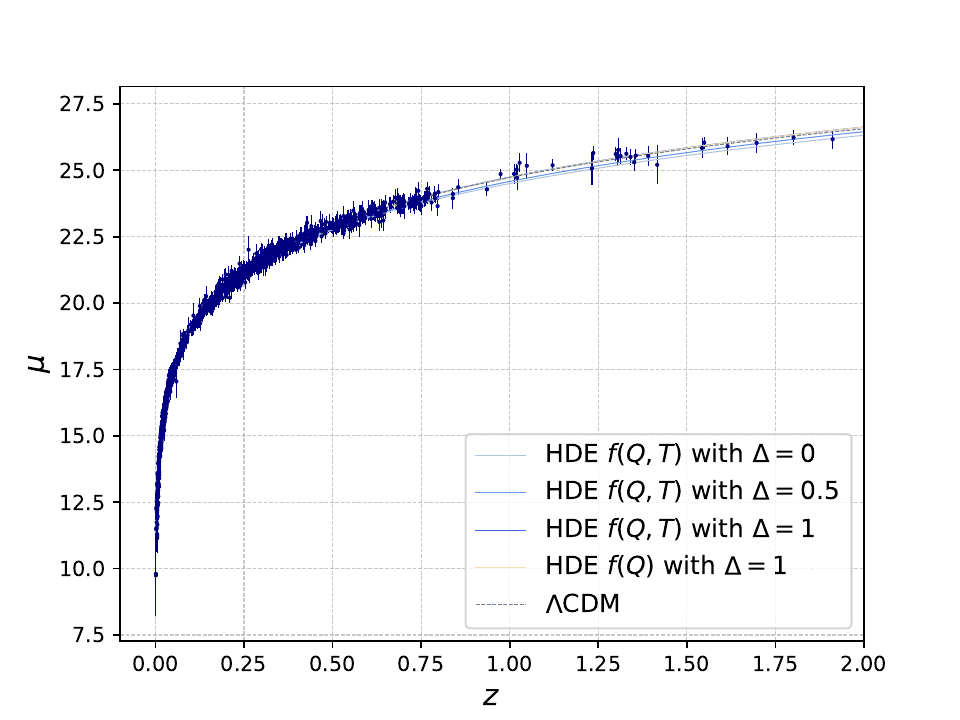}
    \caption{Observational data and best-fit curves for different models: Data points of Cosmic Chronometer (CC) Hubble parameters versus redshift, along with the best-fit curves for each model (left panel). Data points of supernova distance modulus versus redshift along with the best-fit curves for each model (right panel). }
    \label{Fig:ccsnfit}
\end{figure}

\begin{figure}
    \centering
    \includegraphics[width=0.32\textwidth]{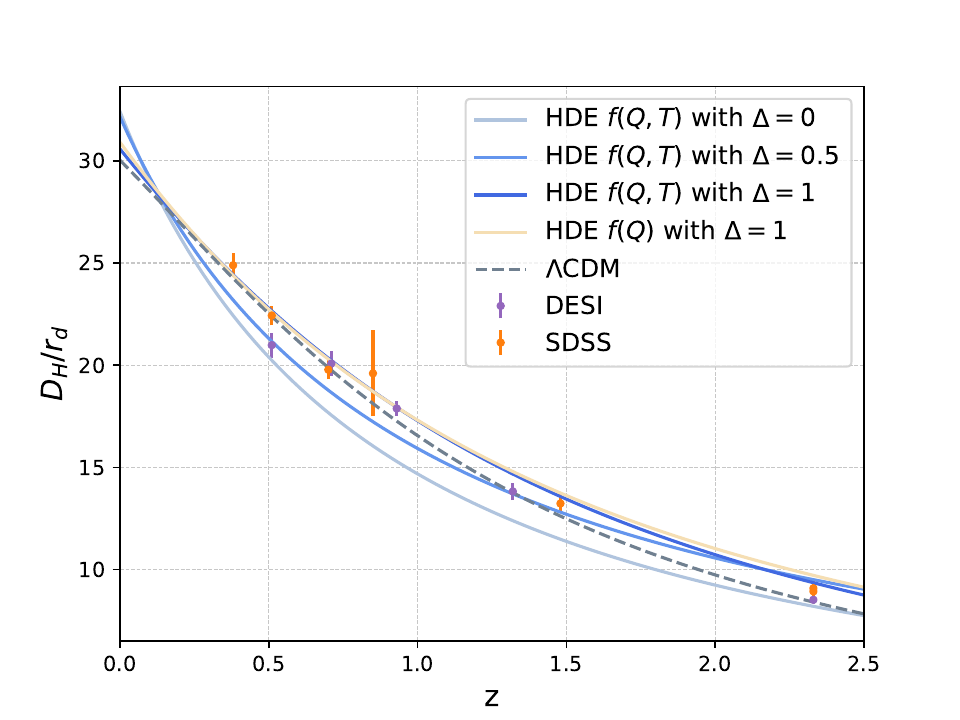}
    \includegraphics[width=0.32\textwidth]{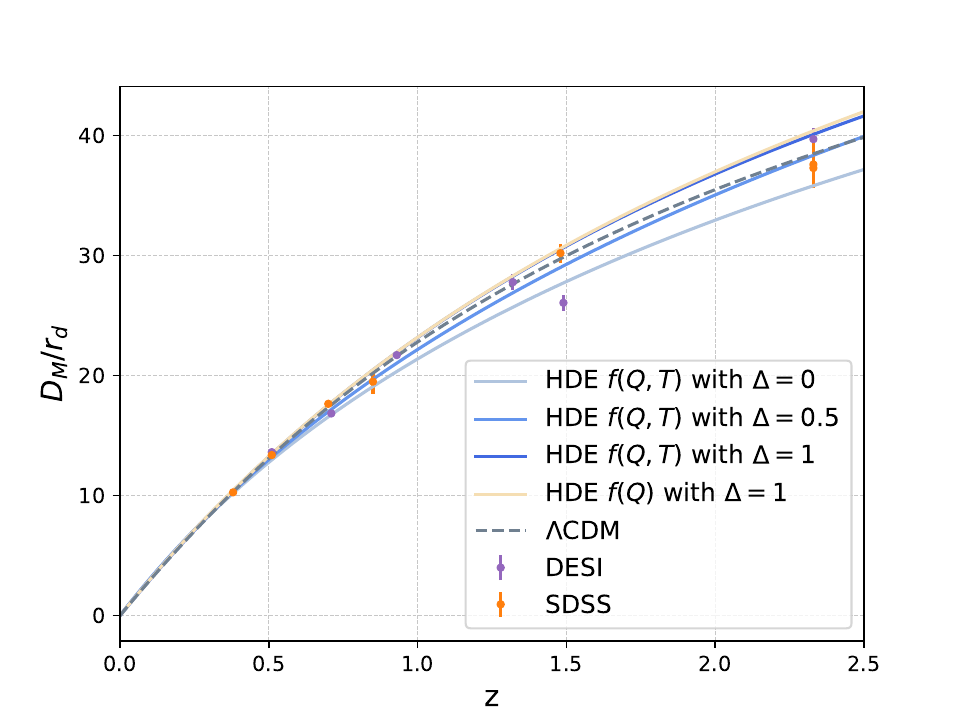}
    \includegraphics[width=0.32\textwidth]{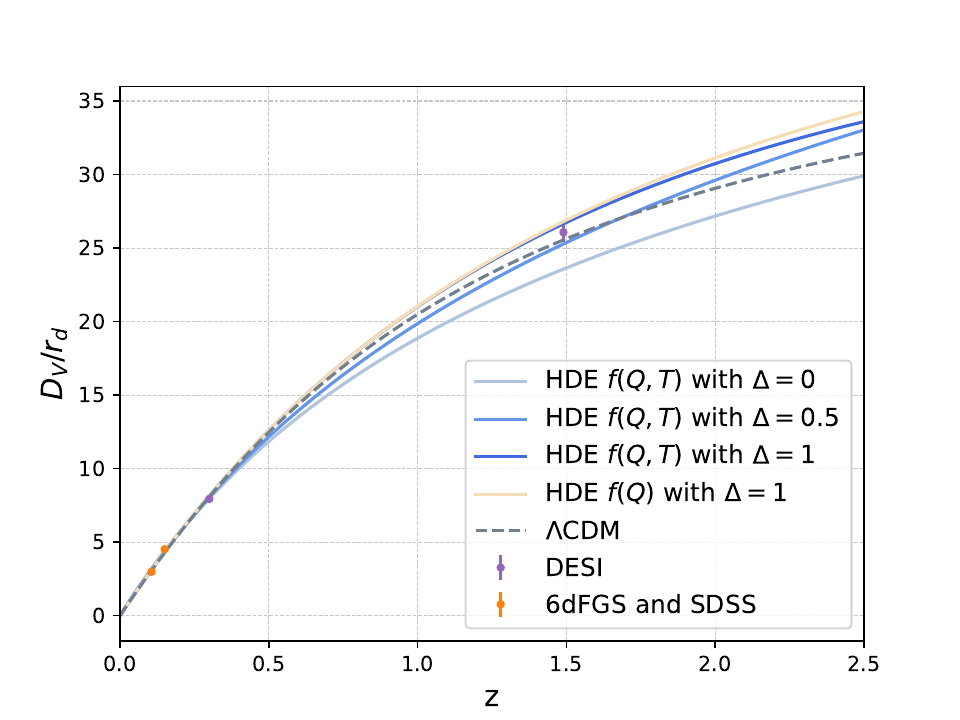}
    \caption{Fitting curves of HDE models in BAO, the error bars represent the data from the 6dFGS, SDSS, and DESI BAO measurements. Hubble distance over the sound horizon at the drag epoch $D_H/r_d(z)$ as a function of redshift $z$ (left panel). The comoving diameter distance over the sound horizon at the drag epoch $D_M/r_d(z)$ as a function of redshift $z$ (mid panel). The angle-average distance over the sound horizon at the drag epoch $D_V/r_d(z)$ as a function of redshift $z$ (right panel).}
    \label{Fig:baofit}
\end{figure}

We first plotted the evolution of the Hubble parameter as a function of redshift under the best-fit scenarios for our 4 different models in left panel of Fig.~\ref{Fig:ccsnfit}. For comparison, we also included the evolution curve of the $\Lambda$CDM model under the same conditions. We find that the larger $\Delta$ parameter presents a better fit and gradually reduces the value of the Hubble parameter at high redshifts. We also observe that there is a divergence in whether to consider non-minimal coupling, with minimal coupling models that ignore the interaction tending to present a flatter curve. This phenomenon suggests that the interaction in the non-minimal coupling model redistributes the energy between matter and dark energy, thereby reducing the dominance of matter at high redshift. We also plotted the relationship between the supernova distance modulus predicted by the model and the redshift  (right panel of Fig.~\ref{Fig:ccsnfit}), along with the data points and error lines obtained from the Pantheon+ dataset, and we found that none of the models were significantly different from the fit of the observed supernova distance modulus. The corresponding best fit curve for BAO data with the data points and error lines are also plotted in  Fig.~\ref{Fig:baofit}.

\begin{table}
    \centering
    \caption{AIC and BIC values for different cosmological models.}
    \begin{tabular}{lcc}
    \hline
    Model & AIC & BIC \\
    \hline
    $\Lambda$CDM & 2058.68 & 2108.09 \\
    $f(Q,T), \Delta=0$ & 2514.35 & 2526.94 \\
    $f(Q,T), \Delta=0.5$ & 2230.17 & 2269.38 \\
    $f(Q,T), \Delta=1$ & 2133.32 & 2172.53 \\
    $f(Q), \Delta=1$ & 2227.35 & 2266.56 \\
    \hline
    \end{tabular}
    \label{tab:AICBIC}
\end{table}

We can evaluate the model by calculating the AIC and BIC as shown in Table \ref{tab:AICBIC}. However, it is to be expected that these models deviate significantly from the standard $\Lambda$CDM model. The reasons for the deviation are not only due to the large differences caused by non-metric gravitational forces, but also the coupling effects between geometric tensors and dynamic tensors, which cannot be ignored. More importantly, these models have little cosmological motivation, and it is impossible to restore them to any particular limit case \cite{rudraObservationalConstraintFRT2021}. Nonetheless, the results also show that when $\Delta=1$, the dark energy confinement is better than the results in other cases, and that the non-minimally coupled case performs better than the minimally coupled case.

\begin{figure}
    \centering
    \includegraphics[width=0.45\textwidth]{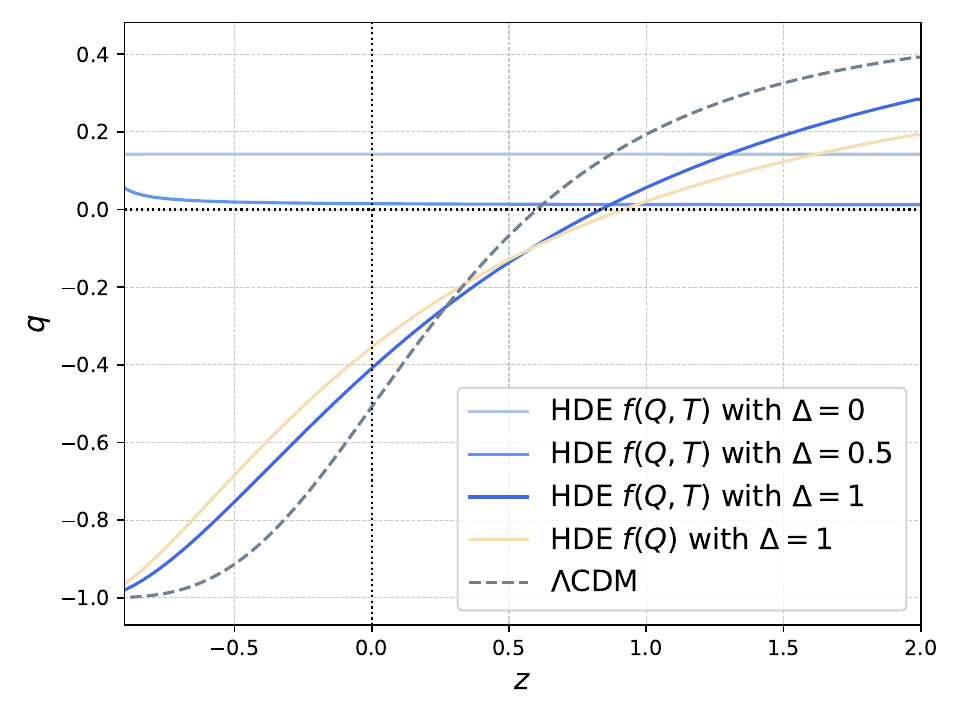}
    \includegraphics[width=0.45\textwidth]{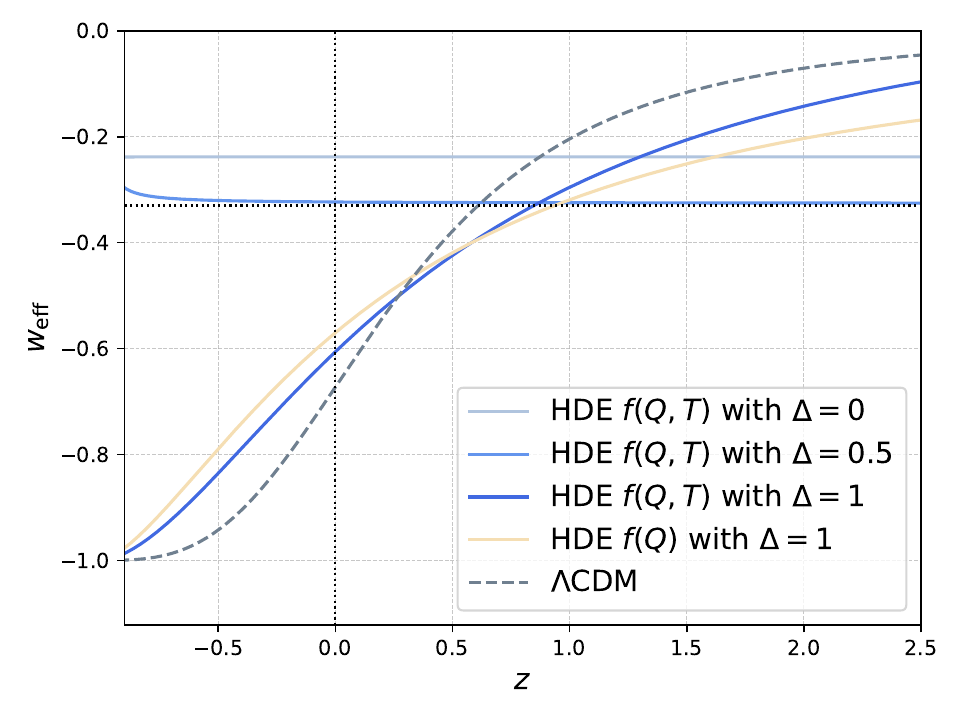}
    \caption{Evolution of the deceleration parameter $q(z)$ (left panel) and the effective equation of state $w_{\text{eff}}(z)$ (right panel) for different models. The curves represent the best-fit trajectories for HDE in $f(Q,T)$ gravity with $\Delta=0, 0.5, 1$, alongside the HDE $f(Q)$ model and the standard $\Lambda$CDM model.}
    \label{Fig:qw}
\end{figure}

\begin{figure}
    \centering
    \includegraphics[width=0.45\textwidth]{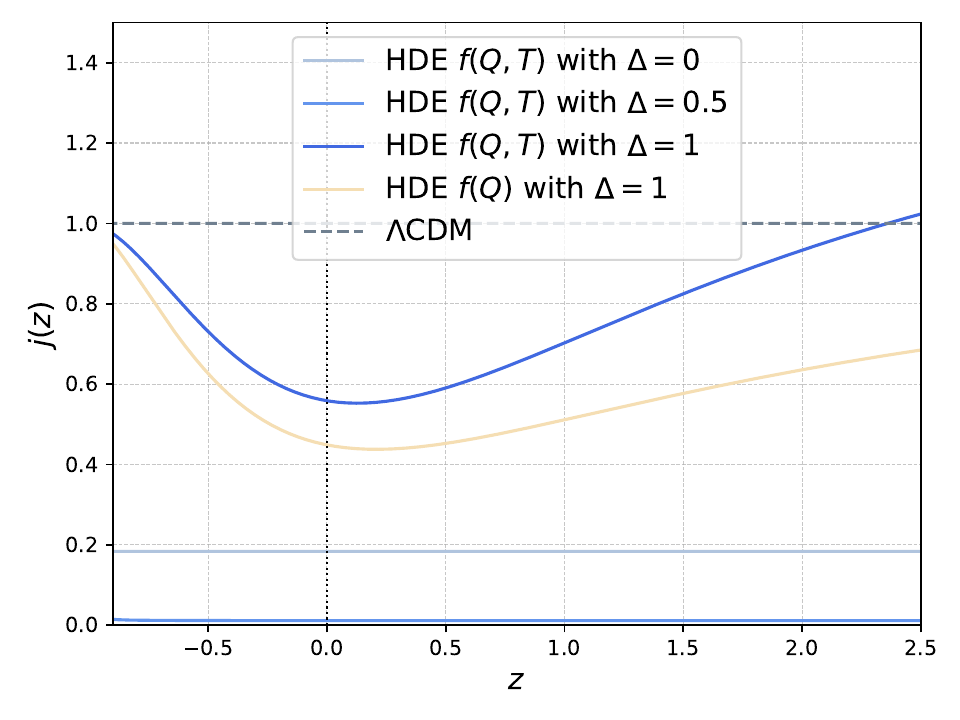}
    \includegraphics[width=0.45\textwidth]{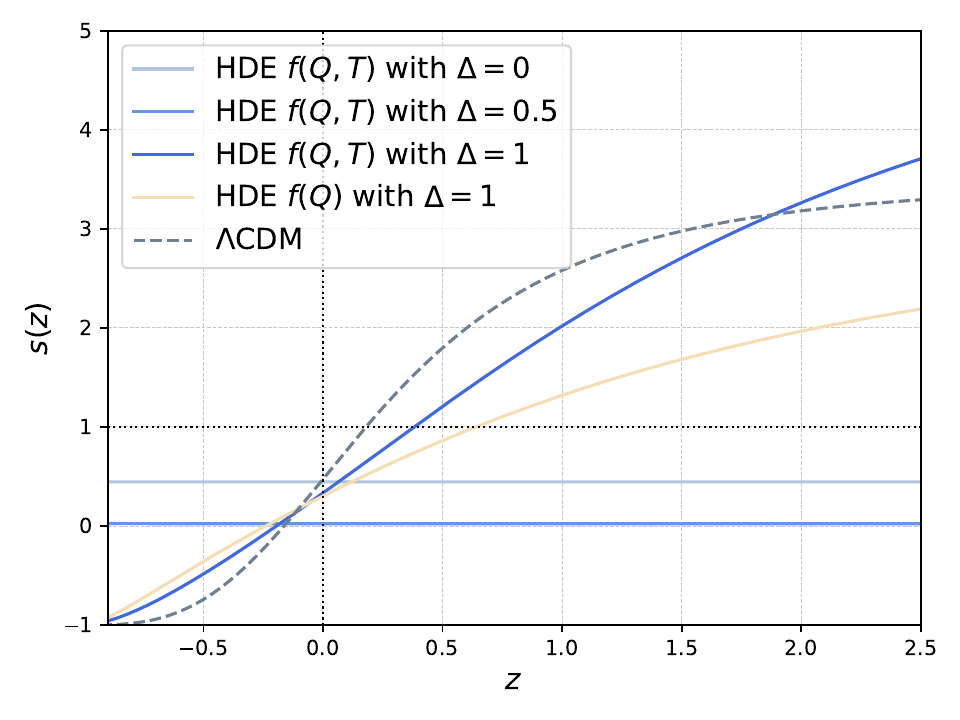}
    \caption{Evolution of the cosmographic parameters: jerk parameter $j(z)$ (left panel) and snap parameter $s(z)$ (right panel). The curves represent the best-fit trajectories for HDE in $f(Q,T)$ gravity with $\Delta=0, 0.5, 1$, alongside the HDE $f(Q)$ model and the standard $\Lambda$CDM model.}
    \label{Fig:js}
\end{figure}

To investigate the dynamic behavior of the universe under the proposed framework, we reconstructed the evolution of the cosmographic parameters: deceleration parameter $q(z)$, jerk parameter $j(z)$, and snap parameter $s(z)$ along with the effective equation of state $w_{\text{eff}}(z)$, as illustrated in Fig. \ref{Fig:qw} and Fig. \ref{Fig:js}.

The evolution of $q(z)$ and $w_{\text{eff}}(z)$ is presented in Fig. \ref{Fig:qw}. For the model with maximal deformation ($\Delta=1$), the deceleration parameter clearly reveals a transition from a decelerated phase ($q>0$) to an accelerated phase ($q<0$) at a transition redshift of $z_t \approx 0.8$, which is in good agreement with current observations. Concurrently, the effective EOS $w_{\text{eff}}$ evolves from a matter-dominated value towards $-1$ in the future, with a present value of $w_{\text{eff}} \approx -0.6$, confirming the driving force of dark energy.
In contrast, for the cases of $\Delta=0$ and $\Delta=0.5$, both $q(z)$ and $w_{\text{eff}}(z)$ appear nearly constant. This behavior arises because, in these specific limits, the analytical solutions reduce to simple power-law expansions, where the scaling of dark energy mimics that of the geometric terms. Consequently, these models lack the dynamical complexity required to describe the cosmic phase transition, suggesting that a significant horizon deformation ($\Delta \approx 1$) is physically necessary to reproduce the observed expansion history.

Furthermore, we examined higher-order cosmographic parameters in Fig. \ref{Fig:js}. The jerk parameter $j(z)$ serves as a sensitive probe to distinguish dynamical dark energy from the cosmological constant. While the standard $\Lambda$CDM model predicts a constant $j=1$, our $f(Q,T)$ model with $\Delta=1$ exhibits a time-varying $j(z)$, indicating a rich dynamical structure in the dark energy component. The deviation to $j \approx 0.5$ near the present epoch ($z \approx 0$) highlights that the universe is currently undergoing a transient dynamical phase, distinguishing our holographic model from the rigid evolution of $\Lambda$CDM. At high redshifts, this indicates that the model correctly recovers the standard matter-dominated era. By comparing the two $\Delta=1$ curves, we observe that the $j(z)$ values for $f(Q,T)$ gravity are consistently slightly larger than those for pure $f(Q)$ gravity. This enhancement can be attributed to the non-minimal coupling between matter and geometry introduced by the $T$-dependence. The interaction term acts as an additional driving force in the field equations, slightly intensifying the rate of change of cosmic acceleration. Similarly, the snap parameter $s(z)$ for the $\Delta=1$ case shows significant evolution, further differentiating it from the $\Delta=0$ and $\Delta=0.5$ scenarios where these parameters remain constant due to the power-law nature of the background evolution.

\section{Conclusion}\label{sec:conclusion}

In this paper, we investigated the evolution of holographic dark energy within the framework of the non-metric modified gravity theory $f(Q,T)$. Assuming the universe is governed by Weyl-Cartan geometry rather than Riemannian geometry, we treated dark energy consistently within this background as a driving force for cosmic expansion. Our approach is motivated by the search for quantum gravity signatures in the late universe. By combining the UV-IR correspondence from the holographic principle with the non-minimal coupling phenomenology inherent in $f(Q,T)$ gravity, we constructed a model that bridges effective field theory bounds with macroscopic cosmic evolution.

We incorporated the generalized Barrow Holographic Dark Energy (BHDE) model to characterize quantum corrections to the horizon entropy, utilizing the Hubble horizon as the most natural infrared cutoff. To obtain analytical insights, we focused on the linear geometric case ($n=1$) and explored three distinct scenarios: the standard Bekenstein-Hawking entropy ($\Delta=0$), maximal deformation ($\Delta=1$), and an intermediate case ($\Delta=0.5$). We specifically analyzed the influence of the non-minimal coupling, which enables energy exchange between the fluid and geometry. It is worth noting that while our results favor $\Delta=1$, this stands in tension with Big Bang Nucleosynthesis (BBN) constraints ($\Delta \lesssim 1.4 \times 10^{-4}$) reported in other studies \cite{BARROW2021136134}, highlighting a discrepancy that warrants further investigation.

To validate the model, we performed a parameter estimation using the latest supernova data, BAO data, and direct measurements of the Hubble parameter via the MCMC method. Our results indicate that the model can effectively alleviate the Hubble constant tension. Specifically, we found a Hubble constant of $H_0 = 67.4^{+1.9}_{-1.9}\, \text{km/s/Mpc}$, which is consistent with Planck 2018 measurements within $1\sigma$, and reduces the tension with local measurements compared to the standard $\Lambda$CDM model. We also observed that the deceleration parameter and the effective equation of state confirm an accelerated expansion phase, consistent with current observations.

Crucially, we validated the physical consistency of our reconstruction approach. Although we did not impose a conserved matter density a priori, the reconstructed matter component exhibits a scaling behavior close to $(1+z)^3$ (with a log-log slope of $\approx 2.75$), confirming that the model successfully recovers the standard matter era dynamically while allowing for non-trivial interactions.

Among the models considered, the AIC and BIC analysis suggests that the maximal deformation case ($\Delta=1$) provides the best fit. Comparison with the minimally coupled case ($\alpha=0$) further suggests that geometry-matter interactions enhance compatibility with observations, justifying the additional complexity introduced by the $f(Q,T)$ framework.

In summary, while this model introduces additional parameters, it yields intriguing mathematical insights and phenomenological results. It distinguishes itself from a simple rescaling of $\Lambda$CDM through the non-conservation of the energy-momentum tensor and the dynamic evolution of cosmographic parameters. This exploration offers potential insights into the role of non-metric gravity in cosmology. Elucidating the nature of dark energy remains an ongoing endeavor, and frameworks like holographic $f(Q,T)$ gravity provide a novel perspective for understanding the interplay between geometry and quantum entropy.

\section*{Acknowledgments}
This work was supported by the National SKA Program of China (Grants Nos. 2022SKA0110200 and 2022SKA0110203).
Our code and chains are available in \url{https://github.com/irosphis/HDE-in-Modified-Gravity}.

% Use the style of numbering in square brackets.
% If nothing is used, default style will be taken.
%\begin{enumerate}[a)]
%\item 
%\item 
%\item 
%\end{enumerate}  

%\begin{itemize}
%\item 
%\item 
%\item 
%\end{itemize}  

%\begin{description}
%\item[]
%\item[] 
%\item[] 
%\end{description}  

% \clearpage %%Remove this from your manuscript

% Uncomment and use as the case may be
%\begin{theorem} 
%\end{theorem}

% Uncomment and use as the case may be
%\begin{lemma} 
%\end{lemma}

\appendix
\section{Derivation of the Hubble Parameter Evolution Equation}
\label{app:derivation}
In this appendix, we provide the derivation of the cosmological evolution equation of $f(Q,T)$ gravity without assuming a specific ansatz for the dark energy equation of state.

We consider the generalized model function:
\begin{equation}
    f(Q,T) = m Q^n + \alpha T,
\end{equation}
where $Q=6H^2$. The partial derivatives are given by:
\begin{equation}
    f_Q = mn Q^{n-1} = mn (6H^2)^{n-1}, \quad f_T = \alpha.
\end{equation}
The time derivative of $f_Q$ is:
\begin{equation}
    \dot{f}_Q = mn(n-1)(6H^2)^{n-2} 12 H \dot{H}.
\end{equation}
Substituting these into the modified Friedmann equation (Eq. \ref{Fr1}):
\begin{equation}
    \rho_{tot} = \frac{f}{2} - 6f_Q H^2 - \frac{2f_T}{1+f_T}(\dot{f}_Q H + f_Q \dot{H}).
\end{equation}
Here, $\rho_{tot} = \rho_m + \rho_\text{de}$. We expand each term as follows:
\begin{align}
    \frac{f}{2} &= \frac{1}{2}m(6H^2)^n - \frac{\alpha}{2}\rho_m, \\
    -6f_Q H^2 &= -6 \left[ mn(6H^2)^{n-1} \right] H^2 = -mn(6H^2)^n.
\end{align}
The term involving the derivatives of the Hubble parameter becomes:
\begin{align}
    \dot{f}_Q H + f_Q \dot{H} &= \left[ mn(n-1)(6H^2)^{n-2} 12H\dot{H} \right] H + mn(6H^2)^{n-1} \dot{H} \nonumber \\
    &= 2mn(n-1)(6H^2)^{n-1} \dot{H} + mn(6H^2)^{n-1} \dot{H} \nonumber \\
    &= mn(2n-1)(6H^2)^{n-1} \dot{H}.
\end{align}
Combining these terms, the Friedmann equation reads:
\begin{equation}
    \rho_m + \rho_\text{de} = m\left(\frac{1}{2}-n\right)(6H^2)^n - \frac{\alpha}{2}\rho_m - \frac{2\alpha mn(2n-1)}{1+\alpha}(6H^2)^{n-1} \dot{H}.
\end{equation}
We rearrange the equation to isolate $\dot{H}$:
\begin{equation}
    \frac{2\alpha mn(2n-1)}{1+\alpha}(6H^2)^{n-1} \dot{H} = - \left[ m\left(n-\frac{1}{2}\right)(6H^2)^n + \rho_\text{de} + \left(1+\frac{\alpha}{2}\right)\rho_m \right].
\end{equation}
Finally, using the relation $\dot{H} = -H(1+z) (dH/dz)$, we obtain the general first-order differential equation for the Hubble parameter:
\begin{equation}
    \frac{dH}{dz} = \frac{1+\alpha}{2\alpha m n (2n-1) (6H^2)^{n-1} H (1+z)} \left[ \rho_\text{de}(H) + \left(1+\frac{\alpha}{2}\right)\rho_m(z) + m\left(n-\frac{1}{2}\right)(6H^2)^n \right].
\end{equation}
This equation governs the dynamics of the universe for any $n$ and any form of $\rho_\text{de}(H)$ (such as the Barrow HDE $\rho_\text{de} \propto H^{2-\Delta}$), incorporating the non-minimal coupling effects explicitly. But it must be solved together with the modified matter continuity equation to get $\rho_m$.


% \bibliographystyle{elsarticle-num}
% \bibliographystyle{cas-model2-names}

% % Loading bibliography database
% \bibliography{cas-refs}

\printbibliography


\end{document}